\documentclass[aps,prl,floatfix,twocolumn,tightenlines,amsmath,amssymb]{revtex4-1}

\usepackage{graphicx}
\usepackage{amssymb}
\usepackage{color}
\usepackage{psfrag}
\usepackage{ifsym}
\usepackage{epstopdf}
\usepackage{soul}
\usepackage{tabularx}

\newcommand{\ket}[1] {| #1 \rangle}

\newcommand{\ii} {\textbf{i}}

\begin{document}

\title{An entangling quantum-logic gate operated with\\ an ultrabright single photon-source}

\author{O. Gazzano$^{1}$, M. P. Almeida$^{2,3}$, A. K. Nowak $^1$, S. L. Portalupi$^1$, A. Lema\^itre$^1$, I. Sagnes$^1$, A. G. White$^{2,3}$ and P. Senellart$^1$}
\affiliation{$^1$Laboratoire de Photonique et de Nanostructures, CNRS, UPR20, Route de Nozay, 91460 Marcoussis, France\\
$^2$Centre for Engineered Quantum Systems \& $^3$Centre for Quantum Computer and Communication Technology, School of Mathematics and Physics, University of Queensland, Brisbane QLD 4072, Australia\\}

\begin{abstract}
\noindent We demonstrate unambiguous entangling operation of a photonic quantum-logic gate driven by an ultrabright solid-state single-photon source. Indistinguishable single photons  emitted by a single semiconductor quantum dot in a micropillar optical cavity are used as target and control qubits. For a source brightness of 0.56 collected photons-per-pulse, the measured truth table has an overlap with the ideal case of $68.4{\pm}0.5$\%, increasing to $73.0{\pm}0.6$\% for a source brightness of 0.17 photons-per-pulse. The gate is entangling: at a source brightness of 0.48, the Bell-state fidelity is above the entangling threshold of 50\%, and reaches  $71.0{\pm}3.6$\% for a source brightness of 0.15.
\end{abstract}

\maketitle
The heart of quantum information processing is entangling separate qubits using multi-qubit gates: the canonical entangling gate is the controlled-\textsc{not} (\textsc{cnot}) gate, which flips the state of a target qubit depending on the state of the control. A universal quantum computer can be built using solely \textsc{cnot} gates and arbitrary local rotations \cite{quantumcomp}, the latter being trivial in photonics. In 2001, Knill, Laflamme and Milburn (KLM) demonstrated that photonic multi-qubit gates, could be implemented using only linear-optical components and projective measurements and feedforward~\cite{KLM}. Since then, many schemes to implement linear-optical \textsc{cnot} gates have been theoretically proposed~\cite{cnot1,cnot2,cnot3} and experimentally demonstrated~\cite{cnot4,cnot5,cnot6,cnot7,cnot8,cnot9,cnot10}. These demonstrations all used parametric down conversion as photon sources, however such sources are not suitable for scalable implementations due to their inherently low source brightness---$10^{-6}$ to $10^{-4}$ photons-per-excitation pulse---and contamination with a small but significant multiple-photon component \cite{weinholdarxiv2008, barbieri, jennewein}.

Semiconductor quantum-dots (QDs) confined in micropillar optical cavities are close to ideal as photon sources, emitting pulses containing one and only one photon, with high efficiency and brightness. QDs have been shown to emit single photons~\cite{michler2001}, indistinguishable photons~\cite{santori2002}, and entangled photon pairs~\cite{gershoni,NJPtoshiba}.   Intrinsically, the dots emit photons isotropically: both tapered single mode waveguides~\cite{claudon2012} and micropillar cavities~\cite{doussenat,nat-comm} have enabled the fabrication of single photon sources with brightness of $\sim$80\%. In the latter case, the Purcell effect further allows reducing the dephasing induced by the solid state environment, yielding photons with a large degree of indistinguishability~\cite{santori2002,varoutsis2005,nat-comm}. 

Very recently, quantum-dot photon sources have been used to drive linear-optical entangling gates: on a semiconductor waveguide chip, where the truth table was measured \cite{toshibaCNOT}; and in bulk polarisation-optics \cite{naturenanoCNOT}, where the gate process fidelity was bounded by measurements in two orthogonal bases \cite{qprocess}. These are necessary, but not sufficient measurements for unambiguously establishing entanglement \cite{White2007}, e.g. a \textsc{cnot} gate has the same truth table as a classical, reversible-\textsc{xor} gate.

Here we show unambiguous operation of an entangling  \textsc{cnot} gate using single photons emitted by a single quantum-dot deterministically coupled to the optical mode of a pillar microcavity.  The source is operated at a remarkably high brightness---above 0.65 collected photons-per-pulse---and successively emitted photons present a mean wave-packet overlap~ \cite{santori2002} between 50$\%$ and 72$\%$. Bell-state fidelities above 50\% are an unimpeachable entanglement witness \cite{White2007}: we see fidelities up to $71.0{\pm}3.6$\%.

\begin{figure}[h!]
\begin{center}
\includegraphics[width=\linewidth]{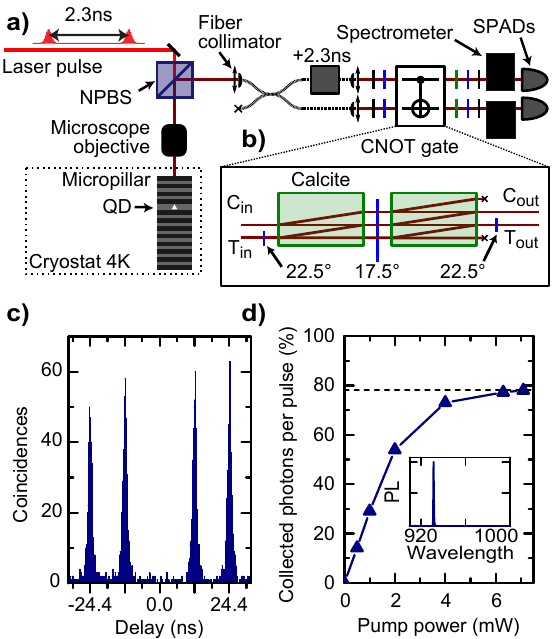}
\caption{a) Schematic of the  the experimental setup. Single photons are produced by a QD in a micro-pillar optical cavity, excited by two consecutive laser pulses, temporally-separated by 2.3~ns. A non-polarising beam splitter reflecting 90\% of the QD signal is employed to send the QD emission into a single-mode fiber and to the input of the CNOT gate. Polarizers, half- and quarter- wave plates are used for state preparation and analysis. The photons are spectrally filtered by two spectrometers and detected by single-photon avalanche photon diodes (SPAD). b) Experimental schematic of the CNOT gate, as described in \cite{cnot4}. c) Autocorrelation function measured on the QD exciton line. d) Collected photons-per-pulse as a function of the pump power. Insert: Emission spectrum of the single photon source. }\vspace{-10mm}
\label{fig:1}
\end{center}
\end{figure}

Our source was grown by molecular beam epitaxy, and consists of an InGaAs annealed QD layer between two Bragg reflectors with 16 (36) pairs for the top (bottom) mirror. After spin-coating the sample with a photoresist, low temperature \emph{in-situ} lithography is used to define pillars deterministically coupled to single QDs~\cite{Dousse2008}. We first select QDs with optimal quantum efficiency and  appropriate  emission wavelength to be spectrally matched to 2.5~$\mu m$ diameter pillar cavities. A green laser beam is used to expose the disk defining the pillar centered on the selected QD with 50~nm accuracy. To operate the source close to maximum brightness and maintain a reasonably high degree of indistinguishability, we use a two color excitation scheme. A  905~nm 82~MHz pulsed laser resonant to an excited state of the QD is used to saturate the QD transition while a low power continuous wave laser at 850~nm is used to fill traps in the QD surrounding, thereby reducing fluctuations of the electrostatic environment. For more details see reference~\cite{nat-comm}. Our source has a maximum brightness of 0.79 photons-per excitation pulse, Fig.~1d, as measured in the first collection lens, Fig.~1a.

The QD emission is collected by a 0.4 NA microscope objective and coupled to a single-mode fiber with a $70\%$ efficiency, estimated by comparing the measured single photon count rate with and without fiber coupling. The typical spectrum of the source is shown in the insert of Fig.~1d, note the single emission line at 930~nm.  To characterize the purity of the single photon emission, we measure the second-order correlation function, g$^2$, using an Hanbury Brown-Twiss setup~\cite{HBT}. Figure 1c. shows the measured auto-correlation function under pulsed excitation only. We obtain g$^2(0){=}0.01{\pm}0.01$---without background correction. For the QD under study, the fine structure splitting of the exciton line is below $2 \mu eV~$\cite{annealingQD}. Thanks to the enhancement of spontaneous emission by the Purcell factor, $F_p{=}3.8$, the photons are indistinguishable in any polarization basis as shown in~\cite{nat-comm} using the Hong-Ou-Mandel experiment. In the following, we operate the source at brightness of 75\% for measuring the gate truth table  and at a brightness of 65\% for demonstrating two-photon entanglement. 

To generate the target and control input photons, the source is excited twice every 12.2ns---the repetition rate of the laser---with a delay between the two excitations of 2.3 ns. The two photons are non-deterministically spatially-separated by coupling the source to a 50/50 fiber beam splitter, and non-deterministically temporally-overlapped by adding the 2.3~ns delay to one of the fibre paths. We implement the \textsc{cnot} gate following the design of reference~\cite{cnot4}, which requires both classical and quantum multi-path interference, Fig.1b. The logical qubits are encoded on the polarization state of the photons with  $\ket{0}{\equiv}\ket{H}$ and  $\ket{1}{\equiv}\ket{V}$. We initialise with polarisers, and set the gate input-state using half-wave plates. Half-wave plates on the control input and output act as Hadamard gates, the internal half-wave plate implements the three 1/3 beamsplitters at the heart of this gate \cite{cnot4}. The waveplates and polarisers on the output modes enable analysis in any polarisation basis. For spectral filtering, the gate outputs are coupled to spectrometers, and are detected via single photon avalanche photodiodes  with 350 ps time resolution. 

\begin{figure*}
\begin{center}
\includegraphics[width=\linewidth]{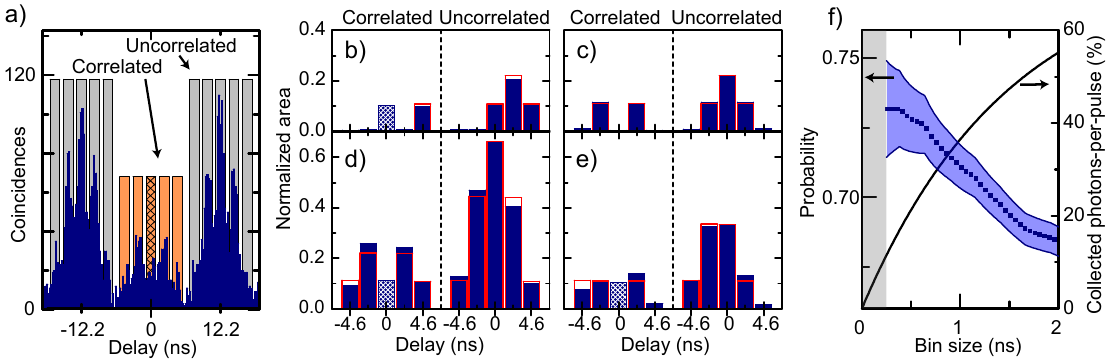}
\caption{(a) Example of a correlation histogram, measured at the output of the \textsc{cnot} gate for input state $\ket{\psi_\text{in}}{=}\ket{1,1}{\equiv}\ket{V,V}$ and output state $\ket{\psi_\text{out}}{=}\ket{1,0}{\equiv}\ket{V,H}$. (b)--(e) Normalized peak areas as a function of time delay or a time bin width of 1ns: correlated peaks at left, uncorrelated peaks at right. For the input $\ket{0,1}{\equiv}\ket{H,V}$, we show the correlation measurements in the basis (b) $\ket{0,1}{\equiv}\ket{H,V}$ and (c) $\ket{1,1}{\equiv}\ket{V,V}$. For the input $\ket{1,0}{\equiv}\ket{V,H}$, we show the correlation measurements in the basis (d) $\ket{1,0}{\equiv}\ket{V,H}$ and (e) $\ket{1,1}{\equiv}\ket{V,V}$.  (f) As a function of time-bin width: \emph{left ordinate} Overlap between measured and ideal truth table for a \textsc{cnot} gate \cite{White2007}, and \emph{right ordinate} Percentage of collected photons-per-pulse \vspace{-7mm}}
\label{fig:2}
\end{center}
\end{figure*}

To obtain the truth table, we measured the output of the gate for each of the four possible logical basis input states $\{\ket{HH},\ket{HV},\ket{VH},\ket{VV}\}$ where $\ket{ct}$ are the control and target qubit states.  Figure 2a presents a typical experimental correlation histogram. Every 12.2~ns, a set of five peaks is observed: each peak corresponds to one of the five possible paths followed by the two photons generated with a 2.3 ns delay. The central peak, at zero delay, corresponds to events where both the control and target photons enter the gate simultaneously. We will hereafter refer to the five central peaks centred at zero delay as \emph{correlated peaks} and the set of peaks centred at $p\times 12.2\ ns$ ($p\in \mathbb{Z^*}$) as \emph{uncorrelated peaks}. For each set of peaks, we also define 5 time bins of variable width, separated by 2.3~ns, in order to temporally analyze the time evolution of the signal. To evaluate the gate properties, we measure the area of the peaks for a given temporal-bin size. Because the emission decay time of the source is 750 ps, adjacent peaks slightly temporally-overlap on the order of 5 to 10\%. The experimental data presented hereafter are corrected for this overlap (see Supplementary information). 

Figures 2b and 2c present the measured area of the correlated and uncorrelated peaks for the control-qubit set to $\ket{0}$, hence the target and control photons do not interfere: the result of the measurement depends only on the purity of the single photon source  $g^2(0)$. In Figures 2d and 2e, the control-qubit is set to $\ket{1}$, and the measurements are obtained with a time bin of 1ns: in this case, the signal measured on the output depends on two-photon interference. For perfectly indistinguishable photons, the peak at zero delay in Fig.~2d should completely vanish, whereas for perfectly distinguishable photons this peak is expected to present the same area as the peaks at $\pm 2.3$~ns. Our observation of an intermediate case highlights the non-unity indistinguishability of successively emitted photons.

Each experimental curve is normalized using the area of the central uncorrelated peaks at $p{\times} 12.2$~ns ($p \in \mathbb{Z^*}$) which can be easily calculated considering the optical path followed  by non-temporally-overlapping photons with Poisson statistics. Doing so, we find that the amplitude of the experimental area (blue bars) averaged over 200 uncorrelated peaks sets is in very good agreement with theoretical expectations (red lines in Figs 2b--e).  This normalization procedure allows us to measure the output coincident count rates normalized to the input pair mode, as shown in Table I for a time-bin width of 1ns. In the 8 logical configurations indicated by $\alpha$ and $\beta$, there is actually no signal on one of the detectors: the dark count to signal ratio leads to $\alpha{<}0.005\frac{1}{9}$ and $\beta{<}0.01\frac{1}{9}$. Using a photon mean-wavepacket-overlap, $M$, of $50\%$ we see that the measured configurations, Table I, are in very good agreement with those predicted for an ideal gate, Table II \cite{cnot2}. The value of $M$ is not corrected for imperfections in the experimental setup---such as visibility of the single photon interference,  polarization ratio of the calcite, etc.---and is therefore a lower bound to the source indistinguishability, and compares well with previously reported values \cite{nat-comm}

\begin{table}[h!]
\begin{center}
\begin{tabularx}{1\linewidth}{|c|XXXX|}
\hline
Input  &	C$_{\ket{\text{HH}}}$		&	C$_{\ket{\text{HV}}}$	&	C$_{\ket{\text{VH}}}$	&	C$_{\ket{\text{VV}}}$	\\
\hline
${\ket{\text{HH}}}$	&$\frac{1}{9}$1.12 	&	$\alpha$			&	$\frac{1}{9}$0.015		&	$\alpha$			\\
${\ket{\text{HV}}}$	&$\alpha$			&	$\frac{1}{9}$0.97	&	$\alpha$				&	$\frac{1}{9}$0.04			\\
${\ket{\text{VH}}}$	&$\beta$			&	$\alpha$			&	$\frac{2}{9}$0.50		&	$\frac{1}{9}$0.92		\\
${\ket{\text{VV}}}$	&$\alpha$			&	$\beta$			&	$\frac{1}{9}$0.75		&	$\frac{2}{9}$0.502		\\
\hline
\end{tabularx}
\caption{Experimental output coincident count rates normalized to the input pair mode. Input $\ket{\psi_\text{in}}{=} \ket{\text{control,target}}$  qubit states are indicated in the left column.\vspace{-5mm}}
\label{Table:1}
\end{center}
\end{table}

\begin{table}[h!]
\begin{center}
\begin{tabularx}{1\linewidth}{|c|XXXX|}
\hline
Input &	C$_{\ket{\text{HH}}}$		&	C$_{\ket{\text{HV}}}$	&		C$_{\ket{\text{VH}}}$	&	C$_{\ket{\text{VV}}}$	\\
\hline
${\ket{\text{HH}}}$			&	$\frac{1}{9}$		&	0		&	0			&	0				\\
${\ket{\text{HV}}}$			&0			&	$\frac{1}{9}$		&	0			&	0				\\		
${\ket{\text{VH}}}$	 		&0			&	0		&	$\frac{2}{9}$(1-M)	&	$\frac{1}{9}$				\\
${\ket{\text{VV}}}$			&0			&	0		&	$\frac{1}{9}$			&	$\frac{2}{9}$(1-M)		\\
\hline
\end{tabularx}
\caption{Theoretical output coincident count rates normalized to the input pair mode. Input $\ket{\psi_\text{in}}{=} \ket{\text{control,target}}$  qubit states are indicated in the left column. \vspace{-7.5mm}}
\label{Table:2}
\end{center}
\end{table}

The left ordinate of Fig.~2f plots the overlap between the measured and ideal \textsc{cnot} gate truth tables---defined as the probability to obtain the correct output averaged over all possible four inputs \cite{White2007}---as a function of time-bin width. The right ordinate of Fig.~2f shows the number of collected photons-per-pulse as a function of the time bin width, given by $I_{max}\times \int_0^{t_{bin}}e^{-t/\tau} dt/\int_0^{\infty}e^{-t/\tau} dt$ where $I_{max}$ is the source operation brightness---here, $I_{max}$=0.75 collected photons-per-pulse---and $\tau{=}750$~ps is the decay time of the single photon emission.  Figure 2f shows that the overlap between the measured and ideal truth table increases from $0.684{\pm}0.005$ for a brightness of 0.56, to $0.730{\pm}0.016$ when reducing the time bin, thanks to improved indistinguishability of photons emitted at shorter delay \cite{varoutsis2005, toshibaPRL2012}.

\begin{figure}[h!]
\begin{center}
\includegraphics[width=0.9\linewidth]{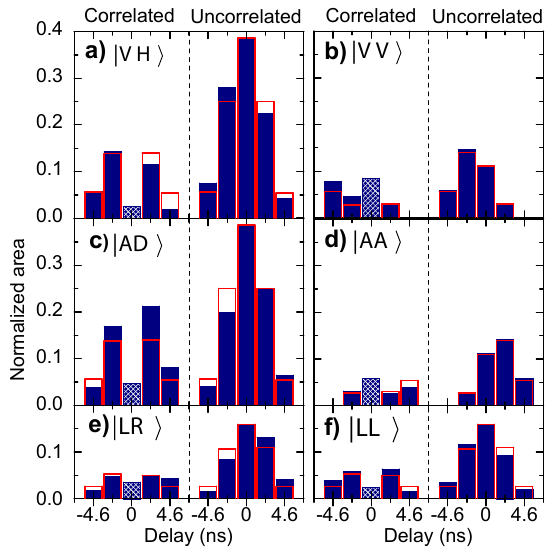}
\caption{a-f) Area of the correlation peaks as a function of time delay for the correlated peaks (left) and uncorrelated peaks (right) for a time bin of 1 ns. For all measurements  the input state is $\ket{D,H}$. The measured output stated are the following: a) $\ket{V,H}$; b) $\ket{V,V}$; c) $\ket{A,D}$; d) $\ket{A,A}$; e) $\ket{L,R}$; and c) $\ket{L,L}$. \vspace{-8mm}}
\label{fig:3}
\end{center}
\end{figure}

To certify that this gate and source combination can produce entangled states from unentangled inputs, we measure the fidelity of the output state with an ideal Bell-state. Setting the control qubit to $\ket{D}{=}(\ket{V}{+}\ket{H})/\sqrt{2}$, and the target qubit to $\ket{H}$, the output of an ideal gate is $\Phi^{+}{=}(\ket{V,V}{+}\ket{H,H})\sqrt{2}$. To measure the fidelity of the experimentally-generated state, we measure the polarization of the correlation in three bases~\cite{White2007,measuringqbit}:
$$ E_{\alpha,\beta}=\frac{A_{\alpha,\alpha}+A_{\beta,\beta}-A_{\alpha,\beta}-A_{\beta,\alpha}}{A_{\alpha,\alpha}+A_{\beta,\beta}+A_{\alpha,\beta}+A_{\beta,\alpha}}$$ where $A_{\beta,\alpha}$ is the zero delay peak area measured for the output control photon detected in $\beta$ polarization and the output target photon in $\alpha$ polarization. The fidelity to the Bell state is then given by $F_{\Phi^{+}}{=}\left(1{+}E_{H,V}{+}E_{D,A}{-}E_{R,L}\right)/4$ where the anti-diagonal polarisation is $\ket{A}{=}(\ket{H}{-}\ket{V})/\sqrt{2}$, and the circular basis polarisations are right, $\ket{R}{=}(\ket{H}{+}\ii \ket{V})/\sqrt{2}$, and left, $\ket{L}{=}(\ket{H}{-}\ii \ket{V})/\sqrt{2}$. Figure 3a-f shows the experimental correlation curves for two polarization configurations in each basis. Note that for both linear and diagonal bases, the results of the measurement depends on the two photon quantum interference only when the output photons are in $\ket{V,H}$, $\ket{V,V}$, $\ket{A,D}$ or $\ket{A,A}$. The four other terms result only from single photon interferences (not shown). 
\begin{figure}[h!]
\begin{center}
\includegraphics[width=1.\linewidth]{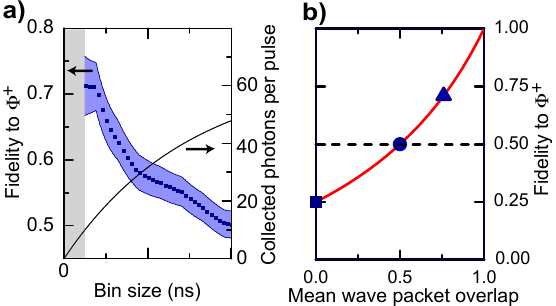}
\caption{a) Fidelity to the Bell State $\Phi^{+}$ and number of collected photons per pulse as a function of time bin. b) \emph{solid line}: Calculated fidelity, F, as a function of the mean wavepacket overlap, M. The symbols correspond to the measured fidelity for uncorrelated peaks (square), correlated zero delay peak with a time bin of 2 ns (circle), and a time bin of 400 ps (triangle). \vspace{-7mm}}
\label{fig:3}
\end{center}
\end{figure}

Figure 4a presents the fidelity to the Bell state $F_{\Phi^{+}}$ as a function of time bin. For all time bins, the fidelity to the Bell state is above the 0.5 limit for quantum correlations.  For these entanglement measurements, we have only slightly decreased the source brightness, $I_{max}$=0.65, in order to obtain a better degree of indistinguishability \cite{nat-comm}. Our results show the creation of an entangled two photon state for a source brightness as large as 0.48 collected photons-per-pulse. When reducing the time bin---and thereby the source brightness, as indicated in the right ordinate of Fig.~4a---the fidelity increases up to $0.710{\pm}0.036$.

 Figure 4b presents the expected fidelity to the Bell state as a function of the mean wavepacket overlap, $M$. Following \cite{cnot2} to calculate the output coincident count rate for all bases configurations, it can be shown that $F_{\Phi^{+}}{=}\frac{1+M}{2(2-M)}$. For $M{=}0$, the fidelity is 0.25, which is the value experimentally observed for the uncorrelated peaks (square). For a time bin of 2 ns, the measured fidelity of 0.5 is consistent with M=0.5 (circle), which is a lower bound for M since our modeling does not take into account the setup experimental imperfections. For a time bin of 400 ps, the measured fidelity of 0.71 shows a mean wavepacket overlap larger than M=0.76 (triangle).

In conclusion, we have demonstrated the successful implementation of an entangling \textsc{cnot} gate operating with an ultrabright single photon source. The gate is entangling for all source brightnesses under 0.48, reaching a Bell-state fidelity of $71.0{\pm}3.6$\% at a source brightness of 0.15 collected photons-per-pulse. To improve the fidelity of the gate operation while maintaining a high source brightness, one could use an adiabatic design of the micropillar to benefit from a larger Purcell effect to further  improve the source indistinguishability~\cite{niels}. The advances on quantum dot single photon technologies open exciting possibilities for linear optical computing. Their main asset as compared to heralded single photon sources based on parametric down conversion is the possibility to obtain very bright sources as well as negligible multiphoton events. Photonic quantum technologies will require access to multiple single-photons, multiplexed in different spatial modes. Small scale implementation of quantum logic circuits is the first step towards incorporating quantum dot based single-photon source to these technologies.

 \begin{acknowledgments}
This work was partially supported by: the French ANR P3N DELIGHT, ANR JCJC MIND, the ERC starting grant 277885 QD-CQED, the French RENATECH network and the CHISTERA project SSQN; and the Australia Research Council Centre for Engineered Quantum Systems (CE110001013) and the Centre for Quantum Computation and Communication Technology (CE110001027). O.G. acknowledges support by the French Delegation Generale de l'Armement; MPA by the Australian Research Council Discovery Early Career Award (DE120101899); and AGW by the University of Queensland Vice-Chancellor's Senior Research Fellowship.
\end{acknowledgments}

\end{document}